\def \AAP #1 #2 {{\em Astron. Astrophys.\/} {\bf #1}, #2}
\def \AAL #1 #2 {{\em Astron. Astrophys. Lett.\/} {\bf #1}, L#2}
\def \AAR #1 #2 {{\em Astron. Astrophys. Rev.\/} {\bf #1}, #2}
\def \AAS #1 #2 {{\em Astron. Astrophys. Suppl. Ser.\/} {\bf #1}, #2}
\def \AJ #1 #2 {{\em Astron. J.\/} {\bf #1}, #2}
\def \ANNREV #1 #2 {{\em Ann. Rev. Astron. Astrophys.\/} {\bf #1}, #2}
\def \APJ #1 #2 {{\em Astrophys. J.\/} {\bf #1}, #2}
\def \APJL #1 #2 {{\em Astrophys. J. Lett.\/} {\bf #1}, L#2}
\def \APJS #1 #2 {{\em Astrophys. J. Suppl.\/} {\bf #1}, #2}
\def \APSS #1 #2 {{\em Astrophys. Space Sci.\/} {\bf #1}, #2}
\def \ASR #1 #2 {{\em Adv. Space Res.\/} {\bf #1}, #2}
\def \BAIC #1 #2 {{\em Bull. Astron. Inst. Czechosl.\/} {\bf #1}, #2}
\def \JSQRT #1 #2 {{\em J. Quant. Spectrosc. Radiat. Transfer\/} {\bf #1}, #2}
\def \MN #1 #2 {{\em Mon. Not. R. Astr. Soc.\/} {\bf #1}, #2}
\def \MEM #1 #2 {{\em Mem. R. Astr. Soc.\/} {\bf #1}, #2}
\def \PLR #1 #2 {{\em Phys. Lett. Rev.\/} {\bf #1}, #2}
\def \PASJ #1 #2 {{\em Publ. Astron. Soc. Japan\/} {\bf #1}, #2}
\def \PASP #1 #2 {{\em Publ. Astr. Soc. Pacific\/} {\bf #1}, #2}
\def \NAT #1 #2 {{\em Nature\/} {\bf #1}, #2}
\def \SAIT #1 #2 {{\em Mem.\ Soc.\ Astron.\ It.\/} {\bf #1}, #2}
\def \MESS #1 #2 {{\em The Messenger\/} {\bf #1}, #2}
\def \ASTRNACH #1 #2 {{\em Astron. Nach.\/} {\bf #1}, #2}
\def\be{\begin{equation}}
\def\ee{\end{equation}}
\def\lsim{\lower 2pt \hbox{$\, \buildrel {\scriptstyle <}\over
         {\scriptstyle \sim}\,$}}
\newcommand\gsim{\buildrel > \over \sim}

\documentstyle{BSAXPwork}
\input epsf.sty
\input psfig.sty
\begin{opening}
\title{Physical Processes in Strong Magnetic Fields of Neutron Stars}
\author{A. K. Harding}
\institute{NASA Goddard Space Flight Center, Greenbelt, MD 20771, USA}
\date{} 
\end{opening}

\begin{document}

\oddpagefooter{}{}{} 
\evenpagefooter{}{}{} 
\medskip  

\begin{abstract}
Neutron stars have inferred surface magnetic fields of up to $10^{14}$ Gauss, in the case of radio pulsars, and up to possibly $10^{15}$ Gauss, in the case of Soft Gamma-Ray Repeaters and Anomalous X-ray Pulsars.  In fields this high, QED effects will profoundly change the characteristics of continuum radiation processes such as synchrotron emission and Compton scattering and will also allow the possibility of additional physical processes such as one-photon pair production, vacuum polarization and photon splitting.  Atomic line processes will also be significantly affected by the presence of strong fields.  I will review some of the properties of radiation processes in strong magnetic fields that are most relevant to pulsars, SGRs and AXPs and the role they play in models for these sources.
\end{abstract}

\medskip

\section{Introduction}

In the last five years there have been a number of neutron stars discovered 
to have magnetic fields of magnitude only imagined by a few 
theorists.  Previous to these recent discoveries, the highest measured neutron star fields 
were around $10^{13}$ G.  The Parkes Multibeam Survey (Manchester et al. 2001, Lorimer 2003) 
has now detected a number of new radio pulsars having surface magnetic fields, 
deduced from their dipole spin-down rates, of near $10^{14}$ G.  Even more remarkable are the Soft Gamma-Ray Repeaters (SGRs) (Kouveliotou 2003) and Anomalous X-Ray Pulsars (AXPs) (Gavril 2003), both now believed to be the predicted magnetars (Thompson \& Duncan 1992) with surface fields possibly as high as $10^{15}$ G.  The fields of these new sources are well above the quantum critical magnetic field of $B_{\rm cr} = m^2c^3/e\hbar$($4.413 \times 10^{13}$ G for electrons), where physical processes are profoundly changed and new processes unique to strong fields dominate.  Observations of these sources can not only provide direct evidence for the high fields but 
will begin to study in detail the exotic processes which can take place only at these field strengths.

This review will discuss the physics of some of the radiative processes taking place in ultrastrong magnetic fields and their possible observational signatures in high-energy wavebands.  Other recent and useful reviews of physics in strong fields have been given by Duncan (1999) and of atomic processes and matter in strong magnetic fields by Lai (2001).

\section{Basics of Physics in Strong Fields} \label{sec:basics}

Magnetic fields can affect radiative processes in a number of ways.  The critical magnetic field strength, $B_{\rm cr} = m^2c^3/e\hbar$ where particle's cyclotron energy equals its rest mass, is a useful measure of when relativistic effects become important.  In neutron star fields, relativistic effects are always important for electrons, whose critical field is $4.413 \times 10^{13}$ G, but never for protons, whose critical field is $1.43 \times 10^{20}$ G. 
Charged particles
occupy Landau states, which arise from the solutions to the Dirac Equation in a homogeneous magnetic field, with energy (in rest mass units) 
\be
E_n  = (1  + p^2  + 2nB')^{1/2} ,  n = l + {\textstyle{1 \over 2}}(s + 1) = 0,1,2, \cdots 
\ee
where $n$, $l$, and $s$ are the principal, orbital and spin quantum numbers, respectively and 
$B' = B/B_{\rm cr}$.  
Particles may have either spin up ($s=1$) or down ($s=-1$), except in the ground state $n=0$, where only the spin-down state is allowed.  The two spin states in levels $n>0$ are thus degenerate.  This degeneracy is however broken by higher order interaction of the particles with the radiation field (Herold et al. 1982, Pavlov et al. 1991) which causes a fine-structure splitting of all the $n>0$ levels.  The momentum component parallel to the field, $p$, is continuous but momentum perpendicular to the field is quantized.  In magnetic fields $B' > 1$, the transverse motion becomes relativistic.  In the non-relativistic limit ($p \ll 1, B' \ll 1$), the transverse energy becomes $nmcB'=neB/mc = n\epsilon_{\rm cr}$, multiples of the cyclotron energy. 

Since the magnetic field can absorb or supply momentum, momentum perpendicular to {\bf B} is not conserved, although parallel momentum and total energy {\it are} strictly conserved.  This allows a number of processes to take place that would be forbidden in free-space.

A magnetic field also strongly affects the photon polarization states and propagation modes.
In a magnetized plasma the two normal modes of propagation are termed the extraordinary (X)
mode, which is mostly polarized perpendicular to the plane containing the field and photon wavevector $\bf k$, and the ordinary (O) mode, which is mostly polarized parallel to the $\bf k-B$
plane (e.g., Meszaros 1992).  The character of these modes is a strong function of the angle between $\bf k$ and $\bf B$, the plasma density and field strength.  In very strong fields, the vacuum electron-positron pairs are polarized (see Section \ref{sec:vac}) and this effect dominates the photon propagation modes, which become almost purely linear (in contrast to the
plasma-dominated modes that are elliptical).

In neutron star magnetic fields, the cyclotron decay rate is high enough so that nearly all 
particles occupy the ground state.  Radiative (rather than collisional) processes thus control the Landau state populations.  This means that resonant scattering (where a photon excites an electron to a higher Landau state from which it then spontaneously decays) dominates over absorption (where a photons excites an electron to a higher Landau state from which it collisionally de-excites).  

\section{Processes Modified by Strong Fields}

\subsection{Cyclotron/Synchrotron radiation}

At low field strengths, cyclotron radiation (and synchrotron radiation at relativistic energies) is a process in which a charged particle radiates as it spirals around the magnetic field.  At high fields, the particle motion perpendicular to the field is quantized and the emission must be described as transitions between Landau states.  When the cyclotron energy, and thus the energies of the emitted photon, is a significant fraction of the particle rest mass (in fields $B \gsim 0.1 B{\rm cr}$), the particle recoil is important.  The classical synchrotron emission formula will violate conservation of energy when the critical frequency, $\epsilon_{\rm cr} = (3/2) \gamma^2 B' \sin\psi$, exceeds the particle's kinetic energy, or when
\be
\gamma ^2 B'\sin \psi  > \left( {\gamma  - 1} \right)
\ee
where $\psi$ is the pitch angle.  Modified formulae (i.e. Sokolov \& Ternov 1968, Harding \& Preece 1987) for synchrotron radiation, based on QED, should be used in this regime.  The quantum synchrotron spectra have a cutoff at the particle kinetic energy (see Figure 1), which limits the emission at high energies.  When $\Gamma \equiv \gamma B'\sin \psi   > 0.1$ there is a decrease in the energy loss rate to $\dot\gamma \propto \gamma^{2/3}$ (Erber et al. 1968), compared to the classical formula where $\dot\gamma \propto \gamma^2$.  A more extreme spectral effect, the dominance of ground-state transitions, occurs for high values of $\Gamma \gsim 1.0$.  In this regime, it is most likely for the particle in an excited Landau state to make a transition directly to the ground state, rather than the classical mode of decaying in a series of many single harmonic number transitions.  This produces a ``tip" at the upper end of the spectrum just before the cutoff, as is evident in the $B' = 10$ case in Figure 1.  Such a feature would be a signature of synchrotron radiation in very high magnetic fields, but would most likely be smeared out by variations in field strength and particle energy in a real source. 
For emission from low Landau states in strong fields, quantum cyclotron transition rates should be used and the spin of the particle is important in determining the spectrum.

\begin{figure} 

\vskip 0.0cm
\epsfysize=6cm 

\hspace{1.0cm} \vspace{0.0cm}
 \epsfbox{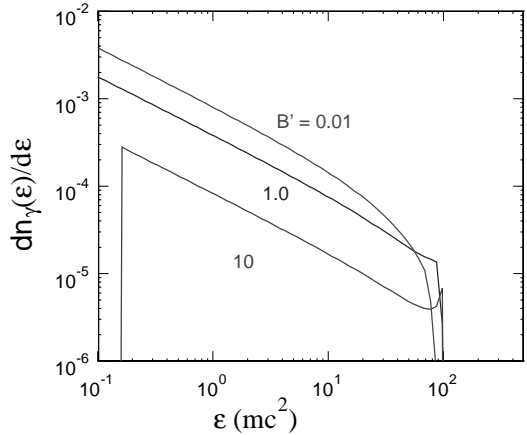}

\caption[h]{Quantum synchrotron spectra for an particle with Lorentz factor $\gamma = 100$, initial pitch angle, $\sin\psi = 1$ and different field strengths $B'$ in units of the critical field.}

\end{figure}

\subsection{Cyclotron absorption and scattering}

Cyclotron absorption, the inverse of cyclotron emission, is a first-order process in which a
photon excites a particle to a higher Landau state.  For particles in the ground state, the 
required energy for excitation to state $n$ is (in rest mass units)
\be
\epsilon _n  = [(1 + 2nB'\sin ^2 \theta )^{1/2}  - 1]/\sin ^2 \theta 
\ee
for a photon propagating at angle $\theta$ to the field (Daugherty \& Ventura 1978).  Because of the recoil of the particle, the cyclotron harmonics are actually anharmonic in high magnetic fields, so that the energy difference between successive harmonics decreases.
As noted in Section \ref{sec:basics}, radiative transitions dominate over collisions in astrophysical sources having strong magnetic fields.  A particle that absorbs a cyclotron photon will almost always de-excite by emitting another photon, rather than collisionally de-exciting, so that the end result of the process is a scattering of the photon rather than a true absorption.  Thus in strong fields cyclotron absorption is most accurately treated as a second-order process, as resonances in the Compton scattering cross section, where the excited state of the particle is a virtual state for which energy and momentum are not strictly conserved.  The cyclotron lines are thus broadened by the intrinsic width which is equal to the inverse of the decay rate from that state (Harding \& Daugherty 1991), although Doppler broadening usually dominates at most angles.  The cross section for Compton scattering in a magnetic field was first studied in the non-relativistic limit by Canuto, Lodenquai \& Ruderman (1971) and the full QED
cross section has been computed by Herold (1979), Daugherty \& Harding (1986) and Bussard, Meszaros \& Alexander (1986).  Since the non-relativistic treatment is limited to dipole radiation, only scattering at the cyclotron fundamental is allowed.  In the relativistic (QED) treatment scattering at higher harmonics is allowed, including Raman scattering, in which the state of the particle after scattering is higher than the initial state.

\subsubsection{Electron cyclotron lines}

When continuum radiation is transmitted through a thermal plasma before reaching an observer,
the radiation at the cyclotron resonant frequencies is scattered most strongly (and out of the line-of-sight), forming a distinctive series of line features.  In most neutron star fields, the electron cyclotron features occur in the hard X-ray region and can be uniquely identified because of their (near) harmonic structure and also because this spectral region is absent of atomic and nuclear lines.  Cyclotron lines are thus an excellent diagnostic tool for providing both a direct {\it in-situ} measurement of magnetic field strength and physical conditions of the radiating plasma.  An electron cyclotron line at 35 keV was first discovered in the accreting pulsar Her X-1 (Trumper et al. 1978) and such lines have since been detected in about a dozen accreting X-ray pulsars (see Orlandini \& dal Fiume 2001, for review).  Magnetic fields have thus been measured for these sources in the range of $10^{12} - 10^{13}$ G.  The most recent and sensitive detections of cyclotron lines have been made by
BeppoSax, which detected four features (presumably the fundamental and three harmonics) in the spectrum of X0115+63 (Santangelo et al. 1999).  Figure 2 shows this detected spectrum along with recent theoretical models of cyclotron resonance lines, illustrating the diagnostic power of such spectral features.  By comparing the observed spectra to even the simplest homogeneous slab and cylinder models of the scattering plasma (where the magnetic field is either perpendicular or parallel to the direction of highest optical depth, respectively) it is possible to identify both the geometry of the plasma (probably slab) and the angle of the observer to the field (about $50^0 - 70^0$).  

\begin{figure}[t] 

\vskip -0.5cm
\epsfysize=8.5cm 

\hspace{-0.8cm} \vspace{0.0cm}
 \epsfbox{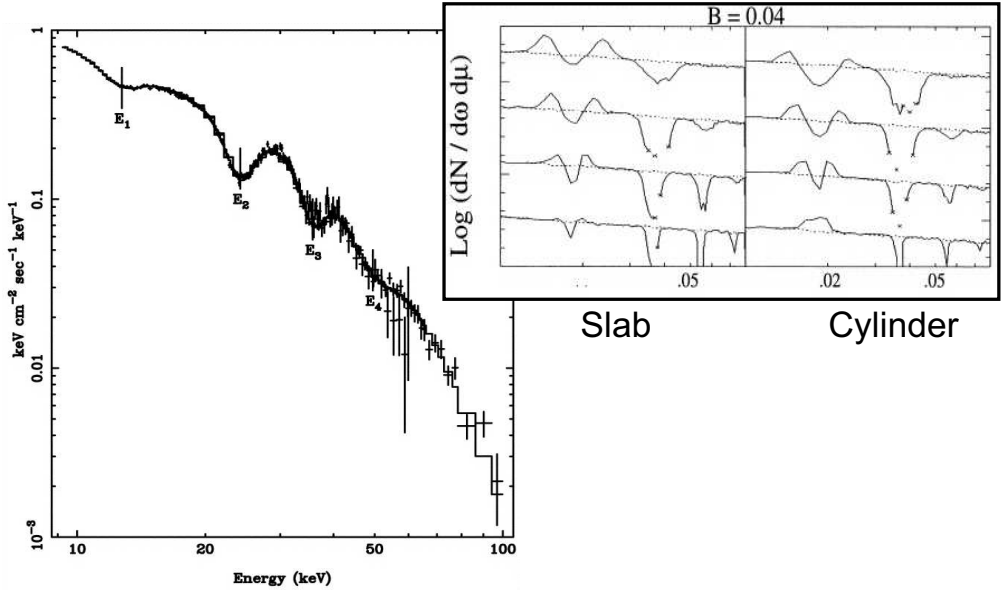}

\caption[h]{Left: Spectrum of X0115+63 (Santangelo et al. 1999) showing four cyclotron features.
Right: Model cylotron resonant scatering features in slab and cylinder geometry for a field strength of $B' = 0.04$, for viewing angles ranges $\cos\theta = >0.75, 0.5-0.75, 0.25-0.5, <0.25$ (top to bottom) (Araya \& Harding 1999)}

\end{figure}

As was noted in Section \ref{sec:basics}, the interaction with the radiation field breaks the spin degeneracy of the excited Landau states, causing a splitting of the levels.  The level splitting is quite small (Pavlov et al. 1991),
\be
\Delta \varepsilon _{sf}  \approx \frac{\alpha }{{2\pi }}B' \approx 10^{ - 3} \varepsilon _{cyc}, 
\ee
$\alpha$ being the fine-structure constant, and produces a fine structure of the cyclotron harmonic features that would be potentially observable with very high resolution.  There are also electron spin-flip transitions possible in the same Landau state which would produce lines at the above energies, in the 1-10 keV range for magnetar fields.  However, the rate of such transitions (Parle 1996, Geprags 1994), $\Gamma _{sf}  \propto \alpha ^6 B'^3$, is extremely small, being a third-order process, so that these lines would not be detectable.

\subsubsection{Proton cyclotron lines}

In the very high fields of SGRs and AXPs, the electron cyclotron fundamental would fall well above 1 MeV and therefore above the observed continuum.  However the proton cyclotron fundamental, $\epsilon_{\rm cyc} = 6.3 (B/10^{15}$G) keV would fall right in the range of the observed spectra.  Proton cyclotron features have been predicted in a number of spectral models (Ho \& Lai 2001, Zane et al. 2001, Ozel 2002, Lai \& Ho 2002).  Very recently, absorption-like features have been detected around 5 keV in the burst spectra of SGR 1806-20 (Ibrabim et al. 2002) which could be proton cyclotron lines if the surface field is really as high as $10^{15}$ G.  A feature in the spectrum of the AXP 1RXS J170849-400910 at around 8 keV has also been reported (Rea et al. 2003), again implying a magnetic field around $10^{15}$ G if the feature is due to proton cyclotron scattering.  One interesting result of the theoretical models of radiation transfer in very high magnetic fields is that vacuum polarization (see Section \ref{sec:vac}) effectively suppresses the strength of the proton cylotron features which would otherwise be much deeper than the observed spectral features. 

\begin{figure}[t] 

\vskip -0.5cm
\epsfysize=6.5cm 

\hspace{-0.8cm} \vspace{0.0cm}
 \epsfbox{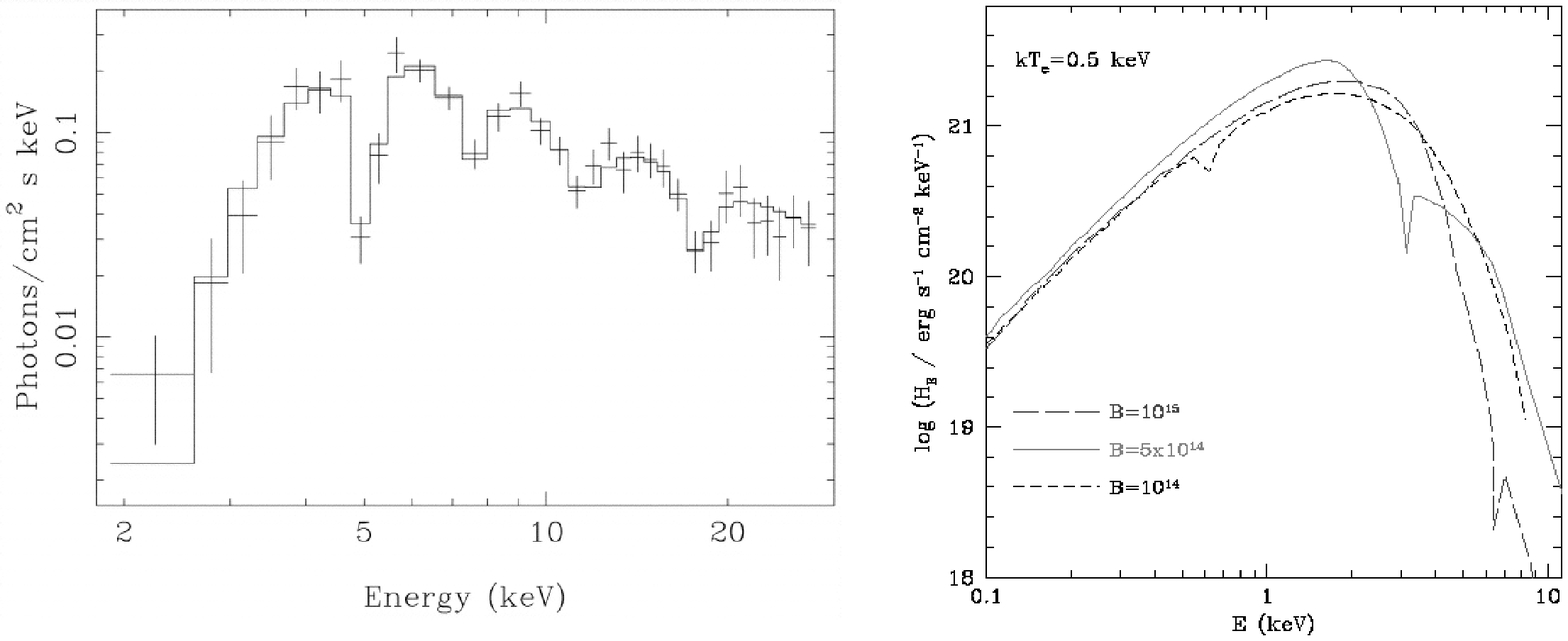}

\caption[h]{Left: Burst spectrum of SGR 1806-20 (Ibrahim et al. 2002) showing possible proton cyclotron feature.
Right: Model spectrum of emission from strongly magnetized neutron star atmospheres (Ozel 2002).}

\end{figure}

\subsection{Atomic processes}

When the Lorentz force on electrons in an atom is comparable to or greater than the Coulomb force,
which occurs for fields
\be
\frac{{B'mc^2 }}{{2Ry}} > 1 \Rightarrow B > 2 \times 10^9 G,
\ee
where $Ry = $ is the Rydberg constant, the atom is distorted into a cylindrical shape and the electron energy levels are closer to Landau states than to atomic states.  Atomic transitions thus become more cyclotron-like.  Among the many interesting consequences (see Wunner \& Herold 1986, for review) are an increase in the binding energy of the atom.  For example, the ionization potential of Hydrogen,
\be
V_o  \cong \frac{{e^2 }}{{\rlap{--} \lambda }}\left( {\frac{B}{{B_{cr} }}} \right)^{1/2}  = 3.8_{} keV_{} \left( {\frac{B}{{B_{cr} }}} \right)^{1/2} 
\ee
moves from the UV into the X-ray range in magnetar fields.  Atoms are thus more tightly bound so that the ionization fraction is lowered by the magnetic field.  Even in the relatively low fields ($\sim 10^8$ G) of some White Dwarfs, atomic emission and absorption lines can provide an accurate measure of the magnetic field of these objects (Angel et al. 1985).

\subsection{Two-photon pair production and annihilation}

The second-order process in which two photons of energy $\epsilon_1$ and $\epsilon_2$ interact to produce an electron-positron pair, and its inverse process, pair annihilation into two photons, are modified in magnetic fields $B \gsim 10^{13}$ G.  Due to the relaxation of momentum conservation perpendicular to the field, the threshold condition (Daughterty \& Bussard 1980) for two-photon pair production becomes
\be
(\epsilon_1\sin\theta_1 + \epsilon_2\sin\theta_2)^2 + 
2\epsilon_1\epsilon_2[1 - \cos(\theta_1-\theta_2)] \ge 4,
\ee
where $\theta_1$ and $\theta_2$ are the angles the photons are propagating with respect to the field. The free-space threshold condition is obtained by dropping the first term on the left-hand side.  It is thus possible for two photons traveling parallel to each other, and also at an angle to the magnetic field, to produce a pair.  The cross section 
is modified by the requirement that each member of the pair must occupy Landau states, producing a series of spikes at the thresholds for each new state combination (Kozlenkov \& Mitrofanov 1987).

The total cross section for the inverse process, two-photon annihilation, begins to decrease around $10^{13}$ G (Bussard \& Daugherty 1980).  As the cross section decreases, the energy of a pair at rest, which is split equally between the two photons at low and zero field strength, are divided unequally so that one of the photons gets almost all of the initial pair energy.  Also at $B > 10^{13}$ G, the cross section for the first order process of one-photon annhilation (see Section \ref{1gamma}) is increasing exponentially with increasing field strength.  
Thus two photon annihilation seems to transform into one-photon annhilation in magnetic fields above $10^{13}$ G.  The spectral lines resulting from two-photon pair annihilation are therefore quite asymmetric in high magnetic fields (Harding \& Baring 1992).  The annihilation line begins to broaden in fields $B' \gsim 0.1$ and at fields $B' > 0.3$ the lines have become nearly a continuum below the kinematic cutoff at 1 MeV.  Thus the characteristic signature of pair annihilation at low fields, a line at 511 keV, is much more difficult to detect at high field strengths.

\section{Processes Unique to Strong Fields}

In addition to the many changes and transformations that occur to well-known processes, strong magnetic fields allow new and exotic processes that occur only in strong field environments.  

\subsection{One-photon pair production and annihilation} \label{1gamma}

The first-order processes of one-photon pair production and its inverse, one-photon pair annihilation, do not take place in field-free environments because energy and momentum cannot be simultaneously conserved.  However, magnetic fields that are strong enough (i.e. near or above $B_{\rm cr}$) can absorb/supply the extra momentum that a photon must have to create or be created by a pair.  The threshold for one-photon pair production is $\epsilon_{1\gamma} = 2/\sin\theta$, where $\theta$ is the angle between the photon momentum and the magnetic field.  Although the threshold is purely kinematic, the attenuation coefficient is exponentially dependent on field strength (Erber 1966) so that the processes effectively does not take place unless $B'\epsilon\sin\theta \gsim 0.2$.  In neutron star fields $B' > 0.1$, the attenuation coeffient is so high that pair production occurs at threshold (Harding \& Daugherty 1983), so that the quantum effects resulting from discreteness of the pair states causes sharp spikes in the attentuation coefficient at thresholds of pair states.  

One-photon pair production has become a critically important process in models of isolated pulsars, where acceleration of particles to energies of 1-10 TeV results in the radiation of photons well above pair threshold in very strong fields.  However, these photons initially have such small angles to the field ($\theta \sim 1/\gamma$, where $\gamma$ is the Lorentz factor of the particle) that they are below threshold and must travel a distance through the curved magnetic field to attain large enough angles.  Since the strength of a dipole (or multipole) field falls quickly as the photon travels away from the neutron star surface, many escape without pair producing.  Observed spectra that originate at radius $r$ near the poles of a neutron star dipole field will thus have high-energy cutoffs at the pair escape energy (Harding et al. 1997, Harding 2002),
\be \label{Ec}
E_c \sim 2\,\,{\rm GeV}\,P^{1/2}\,\left({r\over R}\right)^{1/2}\, {\rm max}
\left\{0.1, \,B_{0,12}^{-1}\,\left({r\over R}\right)^3\right\}
\ee
where $B_{0,12} \equiv B_0/10^{12}$ G, and $P$ and $R$ are the period and radius of the neutron star.  The field strength dependence of the energy of these cutoffs, which may have been detected in the spectra of $\gamma$-ray pulsars (Harding 2001, Thompson 2001), could provide a measure of neutron star magnetic fields and test high-energy pulsar emission models. 

Production of electron-positron pairs has been an important element in pulsar models, since it is believed that pairs may be necessary for coherent radio emission (e.g. Melrose 2000).  Much theoretical work has been devoted to understanding the absence of radio pulsars detected with long periods, defining a ``death line" in $\dot P$-$P$ space, in terms of the inability of very old pulsars to create pairs (e.g. Harding et al. 2002, and references therein). 

\subsection{Vacuum polarization} \label{sec:vac}

Very strong fields are capable of polarizing the vacuum electron-positron pairs.  This affects the photon progagation modes at high frequencies.  The vacuum dominates the propagation modes at photon energies 
\be \label{Vac}
E_V  \gsim 1_{} {\rm{keV}}\left( {\frac{{Y_e \rho }}{{1^{} {\rm{g}}^{} {\rm{cm}}^{ - {\rm{3}}} }}} \right)^{1/2} \left( {\frac{B}{{10^{14} G}}} \right)^{ - 1} 
\ee
(Ventura, Nagel, \& Meszaros 1979) where $Y_e = Z/A$ is the electron fraction and $\rho$ is the plasma density.  Vacuum polarization causes a number of interesting radiative transfer effects with possibly observable signatures.

When the condition of Eqn (\ref{Vac}) above is satisfied, the photon progation modes are determined by the vacuum polarization, where the modes are predominantly linear, rather than by the plasma, where the modes are elliptical.  The photon modes in the magnetized vacuum are thus linear combinations of the plasma modes.  Photons propagating through either density or magnetic field gradients of a neutron star magnetosphere may undergo adiabatic evolution of their normal modes (Heyl \& Shaviv 2000, Lai \& Ho 2002, Ozel 2001, 2002).  For example a photon that is in purely O-mode at relatively high density, where plasma modes dominate, may slowly evolve into the X-mode if it crosses the vacuum resonance and moves into a region where the vacuum modes dominate.  Since the scattering opacities of the O and X modes are very different, this vacuum-induced mode conversion has significant transfer effects for X-ray photons in neutron star, and especially magnetar, magnetospheres.

Another transfer effect caused by vacuum polarization is enhanced mode coupling during scattering (Bulik \& Miller 1997, Ozel 2002).  When photons undergo scattering in a magnetized plasma, they can change their polarization mode, as well as their energy and direction.  When scattering occurs near the vacuum resonance, the partial cross section for scattering into another mode (e.g. Meszaros \& Ventura 1979) is greatly enhanced.  This effect can significantly change the emergent spectrum and angular distribution of radiation from a neutron star by converting photons from modes with high/low opacity into modes with low/high opacity.  Both the effects of adiabatic mode conversion (without scattering) and mode switching (with scattering) have been found to greatly suppress the strength of ion-cyclotron features near the vacuum resonance, by effectively reducing the opacity of the resonant photons. 

\subsection{Photon splitting}

Photon splitting is a third-order process in which a photon splits into two photons of lower energy (Adler 1971).  It cannot occur in field-free environments but the cross section is significant in high fields, where it is able to compete with one-photon pair production.  Because photon splitting has no threshold, high-energy photons propagating at very small angles to the magnetic field in neutron star magnetospheres may split before reaching the threshold for pair production (Harding et al. 1997).  Generally, splitting dominates over pair production in fields $B' \gsim 1.0$.  Thus it may be important in emission models of high-field radio pulsars (Harding et al. 1997) and magnetars (Thompson \& Duncan 1995, Harding \& Baring 1996, Zhang \& Harding 2001).  

\begin{figure}[t] 

\hspace{1.0cm} \vspace{0.0cm}
\psfig{figure=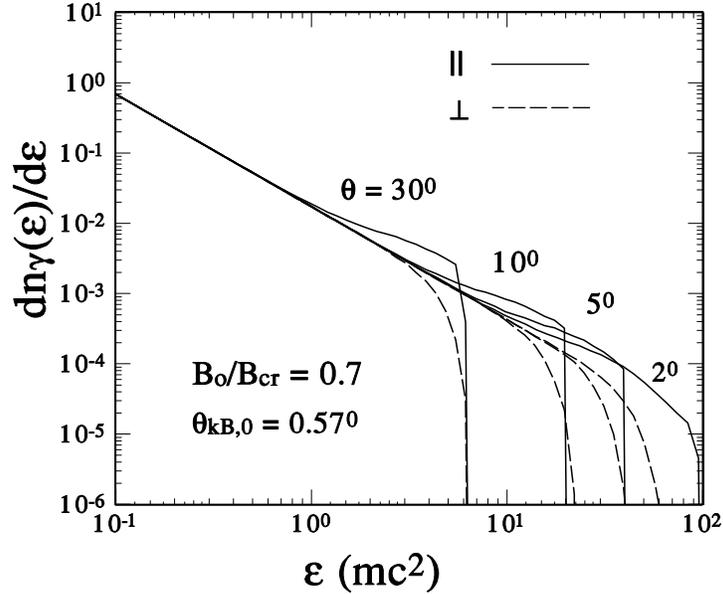,height=9cm,angle=270}

\caption[h]{Model spectrum of emission attenuated by both photon splitting (of $\perp$ mode) and pair production (of $\parallel$ mode) in a neutron star magnetosphere (from Harding et al. 1997)}

\end{figure}

Polarization signatures due to photon splitting may be observable in the spectra of some sources since kinematic selection rules in the low dispersion (low B) limit (Adler 1971) only allows splitting of photons with electric vectors perpendicular to the field (X mode) into two photons with parallel (O) mode polarization.  As a result, parallel-mode photons are attenuated only by pair production while perpendicular-mode photons are attenuated mostly by splitting.  Observed spectra will have polizarization-dependent cutoffs, since the escape energy for photon splitting is lower than the escape energy for pair production, causing the high end of the spectrum to be 100\% polarized in the parallel mode (see Figure 4).

Since photon splitting can dominate, at least in the X-mode, it has the ability to suppress the production of pairs in neutron star magnetospheres.  Baring \& Harding (1998) suggested that the radio-quiescence of SGRs and AXPs could be explained by pair suppression in fields $B' \gsim 1$, if pairs are required for detectable radio emission.  However, complete suppression of pair creation requires that photon splitting be allowed for both photon polarization modes (Baring \& Harding 2001), and thus a break-down of the kinematic selection rules.  But recently, Usov (2002) determined through analysis of the {\it linear} vacuum polarization modes that Adler's kinematic selection rules are valid to arbitrarily high field strengths.  Thus, unless {\it quadratic} vacuum polarization contributions allow a violation of these selection rules, it seems that the parallel mode photons may produce pairs even in magnetar fields and some other reason for their radio quiecsence must emerge.  It is possible that the parallel-mode photons produce bound pairs (Shabad \& Usov 1982) rather than free pairs in high magnetic fields.  If these bound pairs  are stable to dissociation by either the radiation field or a strong electric field (Usov \& Melrose 1995), then complete pair suppression would occur, at least in the strong-field region near the surface.  

While the theoretical jury is still out on whether photon splitting (and bound-pair creation) can cause radio quiescence, the Parkes Multibeam radio pulsar survey (Manchester et al. 2001) is detecting a number of pulsars with very high magnetic fields (as inferred from dipole spin down) near and above the critical field.  Several of these pulsars lie above the photon splitting death line in $\dot P$-$P$  for surface emission (Baring \& Hardng 1998).   However, Zhang \& Harding (2002) have noted that in space-charge limited flow acceleration models, particles can keep accelerating to higher altitudes where fields are low enough for pair production to dominate over photon splitting.  This will move the death line to higher $\dot P$.

\section{Neutron star magnetic fields: Is there a limit?}

The strength of an electric field is  limited by the ``Klein Catastrophe", which occurs when the field is strong enough to cause a breakdown of the vacuum into free electron-positron pairs.  A magnetic field is stable to vacuum breakdown because it cannot accelerate the particles in a vacuum pair (magnetic fields do no work).  But there are several other physical limits that 
apply.   Neutron stars cannot have magnetic fields with pressure high enough to blow them apart.  This gives a limit
\be
\frac{{B^2 R^3 }}{{8\pi }} < \frac{{GM^2 }}{R} \Rightarrow B < 10^{18}\rm G.
\ee
There are also proposed particle physics limits to the strength of neutron star fields.  Various interactions with photons and matter deplete energy and momentum from the field, limiting its strength to $B < 10^{16}-10^{18}$ G  (Lerche \& Shramm 1977).  An ultimate limit of $B < 10^{20}$ G may come from spontaneous production of bound pairs (Zaumen 1976).


\end{document}